\documentclass[prd,twocolumn,aps,amsmath,amssymb,nofootinbib,preprintnumbers]
{revtex4}

\usepackage{graphicx}
\usepackage{dcolumn}
\usepackage{bm}
\usepackage{amsmath}
\usepackage{amsfonts}
\usepackage{epsfig}

\def\ls{\mathrel{\lower4pt\vbox{\lineskip=0pt\baselineskip=0pt
           \hbox{$<$}\hbox{$\sim$}}}}
\def\gs{\mathrel{\lower4pt\vbox{\lineskip=0pt\baselineskip=0pt
           \hbox{$>$}\hbox{$\sim$}}}}
\def\drawbox#1#2{\hrule height#2pt

\hbox{\vrule width#2pt height#1pt \kern#1pt
              \vrule width#2pt}
              \hrule height#2pt}

\def\Asym#1#2{\vcenter{\vbox{\drawbox{#1}{#2}
              \kern-#2pt       
              \drawbox{#1}{#2}}}}

\def\bequ{\begin{equation}}
\def\eequ{\end{equation}}
\def\barr{\begin{array}}
\def\earr{\end{array}}
\def\half{{1\over 2}}
\def\ben{\begin{equation}}
\def\een{\end{equation}}
\def\bena{\begin{eqnarray}}
\def\eena{\end{eqnarray}}

\def\b1{e^0}

\newcommand{\be}{\begin{equation}}
\newcommand{\ee}{\end{equation}}
\def\bea{\begin{eqnarray}}
\def\eea{\end{eqnarray}}

\def\half {{1 \over 2}}

\def\be{\begin{equation}}
\def\ee{\end{equation}}
\def\bea{\begin{eqnarray}}
\def\eea{\end{eqnarray}}

\def\lesssim{\mathrel{\hbox{\rlap{\hbox{\lower4pt\hbox{$\sim$}}}\hbox{$<$}}}}
\def\gtrsim{\mathrel{\hbox{\rlap{\hbox{\lower4pt\hbox{$\sim$}}}\hbox{$>$}}}}

\begin{document}

\title{Emergence of a  Big Bang  singularity in an exact string
background
}

\author{Shinji Hirano$^{1}$}
\author{Anupam Mazumdar$^{2,~1}$}

\affiliation{$^{1}$~Niels Bohr Institute, Blegdamsvej-17, 
Copenhagen-2100, Denmark \\
$^{2}$~Department of Physics, Lancaster University, Lancaster, LA1 4YB, UK}


\begin{abstract}
The origin of Big Bang singularity in $3+1$ dimensions can be 
understood in an exact string theory background obtained by an 
analytic continuation of a cigar like geometry with a nontrivial dilaton.  
In a T-dual conformal field theory picture there exists a closed string 
tachyon potential which excises the singular space-time of a strongly 
coupled regime to ensure that a higher dimensional universe has no 
curvature singularity. However in $3+1$ dimensions the universe 
exhibits all the pathology of a standard Big Bang cosmology. The 
emergence of a singularity  now owes to a higher dimensional 
orbifold singularity which does not have a curvature singularity in 
higher dimensions, suggesting that close to the compactification 
scale an effective description of $3+1$ dimensions 
breaks down and  bouncing universe emerges in $5$ and higher dimensions.
\end{abstract}

\maketitle


For any equation of state obeying the strong energy condition
$p>-\rho/3$, regardless of the geometry (flat, open, closed) of the
universe, the scale factor of the universe in a Friedmann Robertson
Walker (FRW) metric vanishes at $t=0$, and the matter density
diverges. In fact all the curvature invariants, such as $R,~\Box
R,...$, become singular.  This is the reason why it is called the {\it
Big Bang singularity problem}~\cite{Hawking}.

There has been many attempts to resolve this issue by invoking
anisotropic stresses, self regenerating universe (during inflation)
quantum cosmology, etc.~(see~\cite{Linde}) but resolving the  
space-like singularity is particularly hard, especially in the
context of a flat universe~\footnote{In the context of a closed
universe where the curvature term acts as a ``source'' for negative
energy density in the Hubble equation, one can obtain bouncing
solutions~\cite{Starobinsky}.  A particularly interesting proposal of 
a non-singular bouncing cosmology in a flat geometry  has been 
made in~\cite{BMS}, where non-perturbative correction 
to an Einstein Hilbert action leads to an asymptotically free gravity 
and also ghost free. }.

The aim of this paper is to show the {\it emergence} of a Big Bang 
singularity in $3+1$ dimensions within a string theory setup, where 
the scale factor of a flat, homogeneous and isotropic, Friedmann 
Robertson Waker (FRW) metric undergoes a de-accelerating 
expansion, and the scale factor of the universe vanishes in  
finite time. 

However we shall argue that this is an {\it effective}
description of the universe and as we approach near the
compactfication scale (which could be as large as the four dimensional
Planck scale), the Big Bang singular region unfolds to a $4+1$
dimensional world with a {\it bouncing} cosmology. 
The origin of Big Bang singularity is now clear, it is due to a 
{\it boost orbifold singularity} in higher dimensions and 
not as a {\it curvature} singularity.

In order to realize a cosmology which is free from curvature singularity in 
{\it higher dimensions}, we wish to avoid space-like singularity. 
We would also wish to have a background geometry which has 
an exact Conformal Field Theory (CFT) description. In which case 
the space-time is exact in all orders in $\alpha^{\prime}$.  In addition 
if the exact CFT requires a nontrivial dilaton varying in space-time,
then its growth needs to be bounded from above for the quantum 
corrections to be suppressed.

In order to illustrate our setup let us consider a two-dimensional
cigar-like geometry with a space varying
dilaton~\cite{Witten:1991yr,Mandal:1991tz,Elitzur:1991cb}
\bea
\label{cigar_metric}
 ds^2 &=& k\left[ dr^2 + \tanh^2r d\phi^2 \right] \nonumber\\
 \Phi-\Phi_0 &=& -\log\cosh r\,,
\eea
where $\phi$ is a periodic coordinate with $\phi\sim\phi + 2\pi$.  In
string theory the cigar corresponds to an exact conformal field theory
given by the coset $(SL(2,\mathbb C)/SU(2))/U(1)$, where the parameter
$k$ in the metric corresponds to the level of the $SL(2)$ current
algebra.  The central charge is given by $c= 3k/(k-2) - 1$ in the
bosonic case and $c=3(k+2)/k$ in the supersymmetric
case~\cite{Giveon:1999px}.

The metric Eq.~(\ref{cigar_metric}) is exact in the supersymmetric
case and receives ${\cal O}(1/k)$ corrections in the bosonic case. The dilaton
is bounded from above and the string coupling $g_s=e^{\Phi}$ remains
small in the entire space by choosing $e^{\Phi_0}\ll 1$.

Furthermore it has been conjectured that the cigar CFT is equivalent
to the Sine-Liouville model -- FZZ duality~\cite{FZZ,Kazakov:2000pm}
(see also \cite{Fukuda:2001jd}). The sine-Liouville model is defined by
the Lagrangian
\be
 {\cal L} = {1\over 4\pi}\left[(\partial x_1)^2 + (\partial x_2)^2
    + Q\hat{R}x_1 + \lambda \, e^{-x_1/Q}
\cos R x_2'\right],
\label{SL}
\ee 
where we have defined the T-dual coordinate $x_2'\equiv x_{2L}-x_{2R}$
while $x_2=x_{2L}+x_{2R}$.  This is a linear dilaton CFT with
$\Phi-\Phi_0 = -Qx_1$ and the sine-Liouville interaction
\be
T(x_1, x'_2)=\lambda \,e^{-x_1/Q}\cos R x_2'\,,
\ee
depicts a closed string tachyon condensation. The condensate is
exponentially localized, i.e. semi-localized in the $x_1$ direction
and has a winding in the $x_2$ direction.

The parameters of this theory are related to the level $k$ of the
cigar CFT by $Q^2=1/(k-2)$ and $R=\sqrt{k}$.  An analogous conjecture
relates $N=2$ Liouville theory \cite{Kutasov:1990ua} to the
supersymmetric version of the cigar coset CFT
\cite{Hori:2001ax,Tong:2003ik,Eguchi:2004ik,Ahn:2002sx}, in which case
$Q^2=1/k$.  In the Sine-Liouville model $x_2$ is a periodic coordinate
with period $2\pi R$. In the asymptotic weakly-coupled region both the
cigar and Sine-Liouville model look like a cylinder with a linear
dilaton, and the coordinates are identified as $r\sim Qx_1$ and
$\phi\sim x_2/\sqrt{k}$ for large $k$.

\begin{figure}[htbp]
\medskip
\centerline{\epsfxsize=3.5in\epsfbox{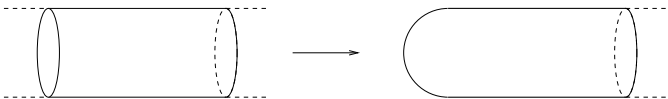}}
\medskip
\caption{Semi-localized tachyon condensation turns the cylinder into 
a cigar.}
\label{cylinder}
\end{figure}

The T-dual sine-Liouville description illustrates a string theory
mechanism of the singularity resolution, see
Fig.\ref{cylinder}~\cite{Bergman:2005qf}. Before the tachyon
$T(x_1,x'_2)$ condenses, i.e. $\lambda=0$, strings can propagate into
$x_1\to-\infty$, where the string coupling $g_s=e^{\Phi}$ blows
up. The space-time has a region of strong coupling singularity. By
condensing the tachyon, i.e. $\lambda\ne 0$, the tachyon wall
$T(x_1,x'_2)$ prevents strings from propagating into the strong
coupling region. The singularity is excised that way. Via FZZ duality
the tachyon condensation manifests itself as a cigar geometry. In this
geometric picture the space-time terminates before the strong coupling
singularity develops.

We wish to apply this mechanism to a cosmological model.  Note,
however, that the cigar geometry itself is not applicable to a real
world cosmology.\footnote{It was shown in \cite{Mizoguchi:2007xa} that gravitons could be localized in a four dimensional submanifold at the tip of the cigar. It was then suggested that the cigar CFT may be useful for the brane-world cosmology.} In order to obtain a temporal dependence in a time
dependent metric, we would require an analytical continuation of the
cigar and the sine-Liouville (SL) CFTs~\cite{Craps:2002ii}:
\be
k\to e^{\pi i}k=-k\ ,\quad 
\left\{\begin{array}{ccc}
r\to -t\ ,& \phi\to -i\varphi &(\mbox{cigar}) \\
x_1\to -i\tau\ ,& x_2\to x_2 &(\mbox{SL})\ .
\end{array}\right.
\ee
The cigar becomes a time-dependent spacetime
\bea
\label{tdepcigar_metric}
 ds^2 &=& k\left[ -dt^2 + \tanh^2t d\varphi^2 \right] \nonumber\\
 \Phi-\Phi_0 &=& -\log\cosh t \ .
\eea
The {\it sine-Liouville} interaction becomes {\it sinh-Liouville} and
the linear dilaton turns into time-like:
\bea
T(\tau, x'_2)&=&\lambda \, e^{-\tau/|Q|}\cosh |R| x_2' \nonumber\\
 \Phi-\Phi_0 &=& -|Q|\tau\ .
\eea
We have now gathered enough ingredients to discuss bouncing cosmology.
We focus on the (world-sheet) supersymmetric case.  The central charge
is $c=3-6/|k|$, or equivalently $\hat{c}=2-4/|k|$.  Furthermore we
will be interested in the case $|k|\gg 1$ so that (i) the curvature is
small and (ii) the central charge $\hat{c}\sim 2$ and the rest of the
space-time can be chosen approximately to be flat, e.g. $\mathbb
R^3\times T^5$ with $\hat{c}_{\mathbb R^3\times T^5}=8$.

Our initial background in the {\it string frame} is $\mathbb
R\times S^1 \times \mathbb R^3\times T^5$ with a time-like linear
dilaton,
\bea
ds^2 &=& -d\tau^2 + dx_2^2 + ds_{\mathbb R^3\times T^5}^2\nonumber\\ 
\Phi - \Phi_0 &=&-|Q|\tau \ ,
\label{LDbackground}
\eea
where $x_2$ is the T-dual of $x'_2$ and $x_2\sim x_2 + 2\pi
\beta$.\footnote{After the analytic continuation, $x'_2=x_{2L}-x_{2R}$
is no longer periodic due to the hyperbolic dependence of the tachyon
condensate $T(\tau,x'_2)$. However, its T-dual $x_2=x_{2L}+x_{2R}$ can
be still periodic by the identification $x_{2L}\sim x_{2L}+\pi \beta$
and $x_{2R}\sim x_{2R}+\pi \beta$ and so will be the angular
coordinate $\varphi$.}

In the Einstein frame ($G_E=e^{-4(\Phi-\Phi_0)/(D-2)}G_S$, where $G_E$
and $G_S$ correspond to Einstein and string frame metrics,
respectively), we have
\bea
ds_E^2 &=& e^{|Q|\tau/2}\left(-d\tau^2 + dx_2^2 + ds_{\mathbb R^3\times 
T^5}^2\right)\nonumber\\
e^\Phi  &=&e^{\Phi_0} e^{-|Q|\tau} \ .
\label{EinLDbackground}
\eea
This is a Milne-type universe.  In the far past $\tau\to -\infty$ the
string coupling blows up and the curvature diverges. Note that the
strong coupling singularity translates to a singularity of the
space-time in the Einstein frame and the singularity is null-type.

We now let the tachyon condense, $T(\tau, x'_2)=\lambda \,
e^{-\tau/|Q|}\cosh |R| x_2' $. Then via FZZ duality~\cite{FZZ,
Kazakov:2000pm} the tachyon condensation is mapped to a change in the
$1+1$ dimensional part of the geometry and the dilaton profile as in
Eq.~(\ref{tdepcigar_metric})~\footnote{A similar idea was explored in
the study of an exact time-dependent string theory
background~\cite{Hikida:2004mp}.}. The metric in the Einstein frame
then yields
\bea
\label{Eincigar_metric}
ds_E^2 &=& \left(\cosh t\right)^{\half}\left[|k|\left( -dt^2 +
\tanh^2t d\varphi^2 \right)+ds_{\mathbb R^3\times T^5}^2\right] \nonumber\\
e^\Phi&=&e^{\Phi_0}(\cosh t)^{-1} \,.  
\eea
The resulting space-time is significantly deformed at early times
$-t\gg 1$ when the tachyon condensation is mostly localized, but
asymptotes to the initial background Eq.~(\ref{EinLDbackground}) at
late times.  Note that there is no longer a strongly-coupled region
nor the curvature singularity as a consequence.  The (null)
singularity at the beginning of the universe is resolved.  To see it,
let us introduce the time $\bar{t}$ by $d\bar{t}=\sqrt{|k|}\left(\cosh
t\right)^{{1\over 4}}dt$. Then the space-time takes the form
\be
\label{FRWcigar_metric}
ds_E^2 = -d\bar{t}^2 +a(\bar{t})^2
ds_{\mathbb R^3\times T^5}^2+ b(\bar{t})^2 d\varphi^2\ .
\ee
The scale factor $a(\bar{t})$ behaves as 
\be
\dot{a}(\bar{t})={1\over 4\sqrt{|k|}}\tanh t\ , \qquad
\ddot{a}(\bar{t})={1\over 4|k|}{1\over (\cosh t)^{9/4}}>0\ .
\ee
The universe is accelerating and the evolution is symmetric under
$\bar{t}\leftrightarrow -\bar{t}$.  At $\bar{t}=0$ the scale factor
$a(\bar{t})= 1$ and $\dot{a}(\bar{t})=0$. The universe bounces from
the contracting to the expanding phase. In the infinite past and
future the acceleration stops and the space-time asymptotes to the
Milne universe.

To summarize, the tachyon condensation has excised the (null) big-bang
singularity and as a consequence rendered the universe of the bouncing
type. Here we wish to make a few remarks.  The scale factor $b(\bar{t})$
shrinks to a zero size linearly in $\bar{t}$ near the bouncing point
$\bar{t}=0$. Since $\varphi$ is periodic, this renders the universe
singular.  The singularity is space-like. However, this is not a
curvature singularity but that of a boost orbifold $\mathbb
R^{1,1}/\mathbb Z$~\cite{Horowitz:1991ap, Khoury:2001bz,
Durin:2005ix}.  The space-time must be properly extended to the \lq\lq
Rindler wedges" to ensure the unitary evolution of string states
propagating through the singularity~\cite{Craps:2002ii}. However,
since our interest is in the states excited only in $\mathbb
R\times\mathbb R^3$ for the purpose of our bouncing cosmology, one may
not concern about the extended Rindler regions.

It is also important to note that there are a few potential sources of
instabilities: (i) The most serious of them is a large back-reaction
due to an infinite blue-shift near the singularity~
\cite{Horowitz:2002mw, Berkooz:2002je, Durin:2005ix,Cornalba:2003kd}. 
The instability was argued to be very severe in the classical gravity 
approximation~\cite{Horowitz:2002mw}. However, this is hardly a 
definitive consensus and a smooth end remains a possibility: The 
winding strings become massless at the singularity, however, these  
states were not taken into account in~\cite{Horowitz:2002mw}. 
Moreover, particles and winding strings are produced near the 
singularity. So it is important to take these effects into account. They 
may conceivably work as agents for smoothing out the singularity~\cite{Durin:2005ix}. 
There is evidence that the Eikonal resummation may render the singularity 
much milder~\cite{Cornalba:2003kd}. So the back-reaction may not be as 
large as it was thought in~\cite{Horowitz:2002mw}. 
(ii) Although propagating tachyons are absent in the type II string,
certain light modes lead to an imaginary part in the one-loop
amplitude due to the asymptotic linear dilaton, signaling a
non-perturbative instability~\cite{Craps:2002ii}. However, since the
string coupling is small in our setup, the decay rate  ${\cal O}(e^{-1/g_s^2})$ 
in this channel is negligible.

Further note that the analytic continuation renders the level $k$ of
the coset CFT negative. This implies that the CFT as a world-sheet
theory is not unitary, reflecting the presence of a time-like
direction. However, the unitarity of our concern is that of the target
space-time theory and it is respected, as mentioned
above~\cite{Craps:2002ii}.

Let us now consider how the metric appears upon compactification.  The
eight dimensional part of the space-time is flat $\mathbb R^3\times
T^5$.  One scenario in our framework is to consider the universe as
$4+1$ dimensions with one extra dimension being small
($\varphi\sim\varphi+2\pi a$ with $a\sqrt{|k|}\ll 1$).

We are then interested in the five dimensional universe compactified on
$T^5$. Upon the dimensional reduction, the radii/scale factors of the
compact space, in effect, provide the additional dilaton coupling to
the Einstein-Hilbert action. In our case the easiest is to perform the
dimensional reduction in the string frame. Since the compact space is
flat, the additional dilaton coupling generated is a {\it constant},
the constant volume of $T^5$. For convenience, we choose it to be
unity. Then the ($4+1$)-dimensional metric in the Einstein frame can
be recast simply from the formula $G_E=e^{-4(\Phi-\Phi_0)/(D-2)}G_S$
with $D=5$ into
\be
\label{5dFRWcigar_metric}
ds_E^2 = \left(\cosh t\right)^{{4\over 3}}\left[|k|\left( -dt^2 + 
\tanh^2t d\varphi^2 \right)+d\vec{x}_3^2\right]\ .
\ee
Introducing the time $\bar{t}$ by $d\bar{t}=\sqrt{|k|}\left(\cosh
t\right)^{{2\over 3}}dt$, the space-time yields the FRW universe with
one extra compact dimension:
\be
\label{FRWcigar_metric}
ds_E^2 = -d\bar{t}^2 +a(\bar{t})^2d\vec{x}_3^2+ b(\bar{t})^2 d\varphi^2\,.
\ee
The velocity and the acceleration of the $3$-spatial part are
respectively given by:
\be
\dot{a}(\bar{t})={2\over 3\sqrt{|k|}}\tanh t\ , \qquad
\ddot{a}(\bar{t})={2\over 3|k|}{1\over (\cosh t)^{8/3}}>0\,.
\ee
The evolution of the universe is qualitatively the same as in the
higher dimensional case, ensuring a non-singular bounce~\footnote{Provided 
that the aforementioned back-reaction is considerably softened, as discussed 
above. Note also that the bounce of \cite{Khoury:2001bz} happens in the fifth 
$S^1$-direction. In contrast, the bounce in our model is with respect to the scale factor 
$a(\bar{t})$ for the flat three dimensional space.}.

If we further compactify the space-time down to $3+1$ dimensions on
$S^1$ in the $\varphi$-direction, the universe becomes
\be
\label{4dFRWcigar_metric}
ds_E^2 = {\sqrt{|k|}\over 2}\left|\sinh 2t \right| \left(-|k|dt^2 +d\vec{x}_3^2\right)\,.
\ee
Introducing the time $\bar{t}$ by $d\bar{t}=dt\,|k|^{3/4}|\half\sinh 2t|^{1/2}$, we have the FRW universe
\be
\label{4dFRWcigar_metric}
ds_E^2 = -d\bar{t}^2 +a(\bar{t})^2d\vec{x}_3^2\,.
\ee
where the velocity and the acceleration of the scale factor is given by:
\begin{eqnarray}
\dot{a}(\bar{t})&=&{1\over \sqrt{|k|}}\coth 2t\,, 
\nonumber \\
\ddot{a}(\bar{t})&=&-{2\sqrt{2}\over |k|^{5/4}}|\sinh 2t|^{-5/2}<0\,.
\end{eqnarray}
In this $3+1$ dimensional effective description, our universe 
mimics a type of hot Big Bang cosmology. There is a Big Bang 
singularity at $\bar{t}=0$, the curvature invariants blow up near
$\bar t=0$ and the decelerating expansion of the universe is: 
$a(\bar{t})\sim\bar{t}^{1/3}$.

However now the origin of Big Bang singularity can be understood 
very well. This is due to an orbifold singularity in higher dimensions. 
To illustrate this let us consider a toy example: If we consider a $5$
dimensional flat space, $ds^2=dr^2+r^2d\phi^2+d\vec{x}_3^2$, and 
compactify it along the $\phi$ circle, we find a similar curvature 
singularity at $r=0$, reflecting the corresponding coordinate singularity 
in five dimensions.

So we interpret the Big Bang singularity as a signature of 
a breakdown of $3+1$ dimensional effective description 
near $\bar{t}=0$. In other words, the universe cannot be viewed 
as $3+1$ dimensional, as one approaches $\bar{t}=0$. It is only 
appropriate to consider the universe as $4+1$ dimensional near 
$\bar{t}=0$, where the universe is {\it free} from curvature singularity.

At this point one might worry about the role of a tachyon in $3+1$ 
dimensions, would it have any cosmological implications.  One should note 
here that $3+1$ dimensional universe and a tachyon description is dual to each 
other, i.e., there is no dynamical role of a tachyon.

Typical of a Big Bang cosmology in $3+1$ dimensions, all the relevant 
problems remain such as flatness, homogeneity and isotropy of
the universe.  Furthermore we have to explain the temperature 
anisotropy of the cosmic microwave background (CMB) 
radiation~\cite{WMAP}. In order to address these issues  the universe
must undergo a phase of cosmic inflation (for a review, see~\cite{anu}).

The cosmic inflation could be triggered  within multiple vacua of 
a string landscape (for a review see~\cite{Land})~\cite{Cliff} and end in an 
observable sector via  Minimally Supersymmetric Standard 
Model (MSSM) inflation~\cite{Maz}, which ensures correct
phenomenology such as observed neutrino masses,  baryon 
number density, and cold dark matter~\cite{AKM}. Otherwise
one could as well resort to a curvaton mechanism to explain the 
observed temperature anisotropy~\cite{Curv}.

To summarize, we have constructed a simple toy model of a hot 
Big Bang cosmology which is embedded in string theory background and
has an exact CFT description. In this model the Big Bang singularity is an 
artifact of higher dimensions, which signals a breakdown of a $3+1$ 
dimensional description of our universe.  Near $\bar t=0$, the correct
description of our universe is  given by a  $4+1$ dimensional bouncing 
cosmology with a boost orbifold. The Big Bang singularity is now manifested 
as an orbifold singularity and not as a curvature singularity. In order to match 
the success of a Big Bang cosmology one would have to introduce a matter 
sector in the geometry, which would then ensure successful inflation 
and its graceful exit.

The complete resolution of a Big Bang singularity now lies in a rather milder 
question; how the boost orbifold instability can be tamed in future. Our setup 
provides a simple stringy framework where these questions can be discussed towards
understanding the origin of our universe.

S.H and A.M are partly supported by 
``UNIVERSENET''(MRTN-CT-2006-035863) and  A. M. is also supported 
by STFC  (PPARC) Grant PP/D000394/1


\end{document}